\documentclass[reqno]{amsart}

\usepackage{type1cm}        
\usepackage{graphicx}        

\usepackage{amsfonts}

\newcommand{\sinc}{\mathrm{sinc}}
\newcommand{\dlift}[1]{{\bf L}_{#1}}
\newcommand{\idlift}[1]{{\bf L}_{#1}^{-1}}
\newcommand{\hold}[1]{{\mathcal H}_{#1}}
\newcommand{\samp}[1]{{\mathcal S}_{#1}}
\newcommand{\real}{\mathbb{R}}
\newcommand{\Rp}{\mathbb{R}_+}
\newcommand{\Z}{\mathbb{Z}}
\newcommand{\C}{{\mathbb C}}
\newcommand{\Err}{{\mathcal E}}
\newcommand{\id}{{\mathrm d}}
\newcommand{\ee}{{\mathrm e}}
\newcommand{\jj}{{\mathrm j}}
\newcommand{\ppi}{{\mathrm \pi}}
\newcommand{\Kopt}{K_{\text{opt}}}
\newtheorem{prop}{Proposition}
\newtheorem{theorem}{Theorem}

\begin{document}

\title[YY Filter]{\Large YY Filter\\ --- A Paradigm of Digital Signal Processing}
\author[M. Nagahara]{Masaaki Nagahara}
\address{
Masaaki Nagahara is with Graduate School of Informatics,
Kyoto University, Kyoto 606-8501, Japan
(nagahara@ieee.org).
}

\maketitle

\begin{abstract}
YY filter, named after the founder Prof.~Yutaka Yamamoto,
is a digital filter designed by sampled-data control theory,
which can optimize the analog performance
of the signal processing system with AD/DA converters.
This article discusses problems in conventional signal processing
and introduces advantages of the YY filter.
\end{abstract}

\section{Introduction}
\label{sec:intro}
YY filter is named after Prof. Yutaka Yamamoto,
who is the founder of the modern sampled-data control theory.
Before introducing the filter, I would like to write about him.

Prof. Yutaka Yamamoto has published a textbook on mathematics
\cite{yybook}
in 1998.
In that year, I was an undergraduate student in 
Kobe University, and I started studying control theory.
I bought the book at that time,
and found it very attractive.
Affected by his book, I desired to be supervised
by Prof. Yutaka Yamamoto in Kyoto University.
I then luckily entered the university, and I began to study
as a graduate student.
I have studied sampled-data control and its application
to digital signal processing.
This study has started by Khargonekar and Yamamoto
\cite{KhaYam96},
which Prof. Yamamoto has been energetically addressing.
Under his supervision, I finished my doctoral thesis
titled 
``{\it Multirate Digital Signal Processing via Sampled-Data $H^\infty$ Optimization},''
\cite{Nag03} in 2003.
This study has been of capital interest to me.
I now begin the introduction of this study, YY filters.

In signal processing, signal reconstruction is a fundamental problem.
For this problem, Shannon sampling theorem \cite{Sha49,Uns00} is
widely used.
This theorem is based on the assumption that 
the analog signal to be reconstructed is fully band-limited
up to the Nyquist frequency.
This assumption is however not realistic, since no real analog signals
are fully band-limited.
To such problems, sampled-data control theory has been applied
in \cite{KhaYam96}.
This is the first article of YY filter, which solves the
delayed signal reconstruction problem.
Based on this study, many researches
have been made:
multirate signal reconstruction \cite{YamNagFuj00},
wavelet expansion \cite{KasYamNag04}
audio signal compression \cite{AshKakNagYam04},
fractional delay filters \cite{NagYam03},
image processing \cite{KakNagKobYam05},
adaptive filtering \cite{NagYam08},
probability density estimation \cite{NagSatYam08},
and repetitive control \cite{NagYam09}.

In this article,
I omit discussion on these applications as space is limited,
and I will concentrate on problems
in Shannon's theorem (or its generalization) 
and advantages of the YY filter
over the conventional theorem.

\section{Problems in Sampling Theorem}
\label{sec:problems}
%
\subsection{Shannon sampling theorem}
\begin{figure}[tb]
\begin{center}
\unitlength 0.1in
\begin{picture}( 28.0000,  4.2000)(  0.0000, -6.0000)
%
\special{pn 8}%
\special{pa 800 400}%
\special{pa 1200 400}%
\special{fp}%
\special{sh 1}%
\special{pa 1200 400}%
\special{pa 1134 380}%
\special{pa 1148 400}%
\special{pa 1134 420}%
\special{pa 1200 400}%
\special{fp}%
%
\special{pn 8}%
\special{pa 1200 200}%
\special{pa 1600 200}%
\special{pa 1600 600}%
\special{pa 1200 600}%
\special{pa 1200 200}%
\special{fp}%
%
\special{pn 8}%
\special{pa 1600 400}%
\special{pa 2000 400}%
\special{dt 0.045}%
\special{sh 1}%
\special{pa 2000 400}%
\special{pa 1934 380}%
\special{pa 1948 400}%
\special{pa 1934 420}%
\special{pa 2000 400}%
\special{fp}%
%
\special{pn 8}%
\special{pa 2000 200}%
\special{pa 2400 200}%
\special{pa 2400 600}%
\special{pa 2000 600}%
\special{pa 2000 200}%
\special{fp}%
%
\special{pn 8}%
\special{pa 2400 400}%
\special{pa 2800 400}%
\special{fp}%
\special{sh 1}%
\special{pa 2800 400}%
\special{pa 2734 380}%
\special{pa 2748 400}%
\special{pa 2734 420}%
\special{pa 2800 400}%
\special{fp}%
\put(14.0000,-4.0000){\makebox(0,0){$\samp{}$}}%
\put(22.0000,-4.0000){\makebox(0,0){$\hold{\phi}$}}%
%
\special{pn 8}%
\special{pa 800 200}%
\special{pa 400 200}%
\special{pa 400 600}%
\special{pa 800 600}%
\special{pa 800 200}%
\special{fp}%
%
\special{pn 8}%
\special{pa 0 400}%
\special{pa 400 400}%
\special{fp}%
\special{sh 1}%
\special{pa 400 400}%
\special{pa 334 380}%
\special{pa 348 400}%
\special{pa 334 420}%
\special{pa 400 400}%
\special{fp}%
\put(0.5000,-3.5000){\makebox(0,0)[lb]{$w(t)$}}%
\put(9.0000,-3.5000){\makebox(0,0)[lb]{$x(t)$}}%
\put(18.0000,-3.5000){\makebox(0,0)[lb]{$c$}}%
\put(24.5000,-3.5000){\makebox(0,0)[lb]{$y(t)$}}%
\put(6.0000,-4.0000){\makebox(0,0){$\phi$}}%
\end{picture}%
\end{center}
\caption{Shannon sampling theorem; the signal $w\in L^2$ is band-limited by $\phi$ and
sampled by the ideal sampler $\samp{~}$.
Then an analog signal $y$ is produced by the hold device $\hold{\phi}$ to reconstruct $x$.}
\label{fig:sampling-theorem}
\end{figure}
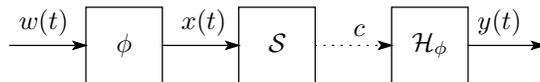

Let $x$  be a continuous-time signal in $L^2$,
the Lebesgue spaces consisting of the square integrable real functions on $\real=(-\infty,\infty)$.
The problem here is to recover the original signal $x$
from its sampled data $\{x(nh)\}_{n\in\Z}$,
where $h>0$ is the sampling period.
This problem is however ill-posed
unless there is an a priori condition on the original signal $x$.
The sampling theorem, usually attributed to Shannon, 
answers this question under the hypothesis of band-limited signals
\cite{Sha49,Uns00}. 
That is, it is assumed that the support of the Fourier transform
$\hat{x}(\jj\omega)$ of $x$ is
limited to the frequency range lower than the Nyquist frequency $\ppi/h$:
\begin{theorem}[Whittaker-Shannon]
Suppose that $x\in L^2$ is fully band-limited, i.e., 
\begin{equation}
x \in BL^2 := \left\{x\in L^2 : \hat{x}(\jj\omega)=0, |\omega|\geq \ppi/h\right\}.
\label{eq:BL-assumption}
\end{equation}
Then the following formula uniquely determines $x$:
\begin{equation}
x(t) = \sum_{n=-\infty}^\infty x(nh)\phi(t-nh),\quad t \in \real,
\label{eq:Shannon-reconstruction}
\end{equation}
where
$\phi(t):= \sinc(t/h) := \sin(\ppi t/h)/(\ppi t/h)$.
\end{theorem}
The reconstruction procedure is shown in Fig.\ \ref{fig:sampling-theorem}.
In this figure, the signal $w\in L^2$ is convoluted (or filtered) by $\phi$, i.e.,
\[
x(t) = \int_{-\infty}^\infty \phi(t-\tau)w(\tau)\id\tau = (\phi\ast w)(t).
\]
Then the signal $x$ is in $BL^2$ (i.e., band-limited) since 
$\hat{\phi}(\jj\omega) = 1$ if $\omega\in(-\ppi,\ppi)$ and $\hat{\phi}(\jj\omega)=0$ if $\omega\notin(-\ppi,\ppi)$.
The signal $x$ is sampled by the ideal sampler $\samp{~}$ with 
the sampling period $h$:
\[
(\samp{~}x)[n] := x(nh),\quad n\in\Z.
\]
Then the discrete-time signal $c=\samp{~}x$ becomes an analog signal $y$ by the hold device $\hold{\phi}$:
\[
(\hold{\phi}c)(t) := \sum_{n=-\infty}^\infty c[n]\phi(t-nh).
\]
By the sampling theorem,
the reconstructed signal $y$ is exactly equal to $x$ (not $w$).

Shannon sampling theorem is a beautiful result 
and is the fundamental theory for the conventional digital signal
processing.
However we can find the following questions in real applications:
\begin{itemize}
\item The band-limiting assumption (\ref{eq:BL-assumption}) 
      does not hold for real signals such as audio, image, or video signals.
\item The reconstruction formula (\ref{eq:Shannon-reconstruction}) 
      is hard to implement on a real device,
      since the sinc function has infinite support, in particular it is not causal.
\end{itemize}
\subsection{Generalized sampling theorem}
The sampling theory mentioned above has been extended
to more general case \cite{UnsAld94,Uns00}, that is,
the function $\phi$ is not necessarily a sinc function,
and the sampler is a generalized sampler $\samp{\phi_1}$ defined by
\[
(\samp{\phi_1}x)[n] := \int_{-\infty}^\infty \phi_1(nh-\tau)x(\tau)\id \tau 
 = \langle x, \phi_1(\cdot-nh)\rangle,\quad n\in\Z. 
\]
In this definition, we have $\samp{\phi_1}x = \samp{~}(\phi_1\ast x)$,
and hence the function $\phi_1$ is considered as the impulse response of the acquisition device.
\begin{figure}[tb]
\begin{center}
\unitlength 0.1in
\begin{picture}( 24.7000,  4.5000)(  7.2500, -6.0000)
%
\special{pn 8}%
\special{pa 1200 200}%
\special{pa 1600 200}%
\special{pa 1600 600}%
\special{pa 1200 600}%
\special{pa 1200 200}%
\special{fp}%
%
\special{pn 8}%
\special{pa 2396 200}%
\special{pa 2796 200}%
\special{pa 2796 600}%
\special{pa 2396 600}%
\special{pa 2396 200}%
\special{fp}%
%
\special{pn 8}%
\special{pa 2796 400}%
\special{pa 3196 400}%
\special{fp}%
\special{sh 1}%
\special{pa 3196 400}%
\special{pa 3128 380}%
\special{pa 3142 400}%
\special{pa 3128 420}%
\special{pa 3196 400}%
\special{fp}%
\put(14.0000,-4.0000){\makebox(0,0){$\samp{\phi_1}$}}%
\put(25.9500,-4.0000){\makebox(0,0){$\hold{\phi_2}$}}%
%
\special{pn 8}%
\special{pa 800 400}%
\special{pa 1200 400}%
\special{fp}%
\special{sh 1}%
\special{pa 1200 400}%
\special{pa 1134 380}%
\special{pa 1148 400}%
\special{pa 1134 420}%
\special{pa 1200 400}%
\special{fp}%
\put(8.5000,-3.2000){\makebox(0,0)[lb]{$x(t)$}}%
\put(16.5000,-3.2000){\makebox(0,0)[lb]{$c_1$}}%
\put(29.5000,-3.2000){\makebox(0,0)[lb]{$y(t)$}}%
%
\special{pn 8}%
\special{pa 1800 200}%
\special{pa 2200 200}%
\special{pa 2200 600}%
\special{pa 1800 600}%
\special{pa 1800 200}%
\special{fp}%
\put(20.0000,-4.0000){\makebox(0,0){$K$}}%
%
\special{pn 8}%
\special{pa 2200 380}%
\special{pa 2400 380}%
\special{dt 0.045}%
\special{sh 1}%
\special{pa 2400 380}%
\special{pa 2334 360}%
\special{pa 2348 380}%
\special{pa 2334 400}%
\special{pa 2400 380}%
\special{fp}%
%
\special{pn 8}%
\special{pa 1600 380}%
\special{pa 1800 380}%
\special{dt 0.045}%
\special{sh 1}%
\special{pa 1800 380}%
\special{pa 1734 360}%
\special{pa 1748 380}%
\special{pa 1734 400}%
\special{pa 1800 380}%
\special{fp}%
\put(22.5000,-3.2000){\makebox(0,0)[lb]{$c_2$}}%
\end{picture}%
\end{center}
\caption{Generalized sampling theorem; the signal $x\in V(\phi_2)$ is
sampled by the generalized sampler $\samp{\phi_1}$
to become the discrete-time signal $c_1$.
Then $c_1$ is filtered by $K$ to become $c_2$.
Finally, an analog signal $y$ is produced by the hold device $\hold{\phi}$ to reconstruct $x$.}
\label{fig:gen-sampling-theorem}
\end{figure}
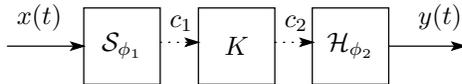
Fig.\ \ref{fig:gen-sampling-theorem} shows a generalized situation.
In this figure, the analog input $x$ is sampled by the generalized sampler $\samp{\phi_1}$.
Then the sampled signal $c_1$ is filtered by a digital filter $K$, and then 
an analog signal $y$ is obtained by the hold device $\hold{\phi_2}$.
In this setting, a generalized sampling theorem is proposed by \cite{UnsAld94}.
The idea is the notion of {\it consistency}:
the output $y$ in Fig.\ \ref{fig:gen-sampling-theorem} can be perfectly
reconstructed by the same system, that is,
for all $n\in\Z$,
\[
\langle x,\phi_1(\cdot-nh)\rangle = \langle y,\phi_1(\cdot-nh)\rangle.
\]
This implies that the reconstruction system works as a projector.
To achieve consistency, the optimal filter $K$ which is linear and time-invariant (LTI)
is constructed by the {\it oblique projection} of $x$ onto $V(\phi_2)$
perpendicular to $V(\phi_1)$,
where $V(\phi_1)$ and $V(\phi_2)$ are closed subspaces in $L^2$, which is defined by
\[
V(\phi_i):=\left\{x=\sum_{n=-\infty}^\infty c[n]\phi_i(t-nh), c \in
\ell^2\right\},\quad i=1,2.
\]
By the oblique projection, the following
generalized sampling theorem is obtained \cite{UnsAld94}.
\begin{theorem}[Unser and Aldroubi]
Suppose that $x\in V(\phi_2)$ and the filter 
\[
A_{12}(z) = \sum_{n=-\infty}^\infty \langle \phi_1(\cdot-nh),\phi_2 \rangle z^{-n}
\]
is invertible\footnote{
This means that $A_{12}(z)$ has no zeros on the unit circle in $\C$.}.
Then the following formula uniquely determines $x$:
\[
x(t) = \sum_{n=-\infty}^\infty (c_1\ast k)[n]\phi_2(t-nh),\quad t\in\real
\]
where $c_1:=\samp{\phi_1}x$ and $k$ is the impulse response of 
$K(z) = A_{12}(z)^{-1}$.
\end{theorem}
The assumption $x\in V(\phi_2)$ can be interpreted as a {\it generalized band-limiting} condition.
Then we again have a problem of non-band-limited inputs, that is, $x\notin V(\phi_2)$.
In this case, the reconstructed signal $y$ can have a large error \cite{NagYamKha08}.
Moreover, it is possible that the optimal filter will be unstable.
This problem is discussed precisely in the next subsection.
\subsection{Causality and stability}
In real-time systems, {\it causality} is a necessary condition for signal processing.
For the sake of simplicity,
we assume \footnote{
If the assumption does not hold, the reconstruction system can be non-causal.}
that $\phi_1(t)=\phi_2(t)=0$ if $t<0$.
Then the causality of the reconstruction system in Fig.\ \ref{fig:gen-sampling-theorem}
depends on the causality of the filter $K$.
If the impulse response $\{k[n]\}$ of the filter $K$ satisfies $k[n]=0$, $n<0$,
then the reconstruction system is causal.
However, in many cases, the filter $K$ may be non-causal, for example,
in the case of polynomial splines \cite{UnsAldEde93b} and exponential splines \cite{UnsBlu05,Uns05}.
This is because the filter $K(z)$ has poles outside of the unit circle in $\C$
\cite{UnsAldEde93b,NagYamKha08}.
In particular, 
it is shown \cite{NagYamKha08} that
high order exponential splines can produce filters
with poles outside the unit circle
provided that the sampling time is sufficiently small.
Therefore, if the non-causal filter is realized as a causal one, 
the poles outside the unit circle lead to an {\it unstable} filter.

\subsection{Summary}
The problems in (generalized) sampling theorem discussed
above are the following:
\begin{enumerate}
\item If the input signal $x$ is not (generalized) band limited, 
the reconstructed signal can show a large error.
In other words, the reconstruction is {\it not robust} against
uncertainty of input signals.
\item The reconstruction system can be non causal.
\item The causal realization of the reconstruction filter $K$ can be unstable.
\end{enumerate}

\section{Sampled-data $H^\infty$ Optimal Signal Reconstruction --- YY Filter}
As we see in the previous section, (generalized) sampling theorem has three problems:
robustness, causality and stability.
In this section, we introduce a new signal processing, 
{\it sampled-data signal processing},
or {\it YY filter},
which is based on sampled-data control theory.
\subsection{Problem formulation}
The main reason to adopt sampled-data control theory is that
we can design a digital filter which optimizes the {\it intersample behavior}.
In other words, we can minimize the reconstruction error for non-band-limited signals.
Moreover, we adopt the {\it $H^\infty$ performance index} for this optimization.
By $H^\infty$ optimization, we can gain the robustness against the input uncertainty.

In the sampling theorem, the optimal reconstruction is a projector on a subspace in $L^2$,
in which for every input the error is minimized in $L^2$ sense.
This means that the error depends on the input and there can be an input for which the error can be arbitrary large.
On the other hand, $H^\infty$ optimization is an optimization for the 
{\it worst case},
by which we can guarantee an error level for {\it any} inputs.
This leads to the robustness against the input uncertainty.

It is obvious that there is no optimal filter $K$ which minimizes the error for {\it all} signals in $L^2$,
or the optimal filter can be $K=0$.
To reconstruct or interpolate the intersample data,
we should assume some {\it a priori} information for the inputs.
Therefore, we assume that the inputs are in the following subspace in $L^2$,
\[
FL^2 := \{Fw: w \in L^2(\Rp)\}
\]
where $F$ is an analog filter which is stable and strictly causal,
and $L^2(\Rp)$ is the Lebesgue spaces consisting of the square integrable real functions on $\Rp=[0,\infty)$.
The filter $F$ is an analog model of the input signals.
The space $L^2(\Rp)$ is a subspace of $L^2$, by which
we can take account of causality and stability of the reconstruction system.
Our signal subspace $FL^2$ is in a sense larger than $BL^2$ or $V(\phi_2)$
because every signal in $BL^2$ or $V(\phi_2)$ can be expanded by $\{\sinc(t-nT)\}$
or $\{\phi_2(t-nT)\}$,
on the other hand, $FL^2$ needs $\{\phi(2^{-m}(t-nT))\}$ for some $\phi$
(wavelet expansion \cite{VetKov}).
In other words, a signal in $FL^2$ can contain arbitrary high frequency components,
the decay rate of which is governed by the filter $F$.

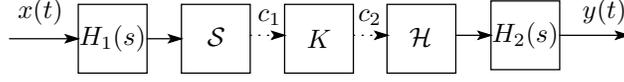
\begin{figure}[tbp]
\begin{center}
\unitlength 0.1in
\begin{picture}( 32.4000,  4.2200)(  2.0000, -5.5500)
%
\special{pn 8}%
\special{pa 1100 196}%
\special{pa 1460 196}%
\special{pa 1460 556}%
\special{pa 1100 556}%
\special{pa 1100 196}%
\special{fp}%
%
\special{pn 8}%
\special{pa 2176 196}%
\special{pa 2536 196}%
\special{pa 2536 556}%
\special{pa 2176 556}%
\special{pa 2176 196}%
\special{fp}%
\put(12.8000,-3.7500){\makebox(0,0){$\samp{}$}}%
\put(23.5600,-3.7500){\makebox(0,0){$\hold{}$}}%
%
\special{pn 8}%
\special{pa 920 196}%
\special{pa 560 196}%
\special{pa 560 556}%
\special{pa 920 556}%
\special{pa 920 196}%
\special{fp}%
%
\special{pn 8}%
\special{pa 200 376}%
\special{pa 560 376}%
\special{fp}%
\special{sh 1}%
\special{pa 560 376}%
\special{pa 494 356}%
\special{pa 508 376}%
\special{pa 494 396}%
\special{pa 560 376}%
\special{fp}%
\put(2.4500,-3.0300){\makebox(0,0)[lb]{$x(t)$}}%
\put(15.0000,-3.0300){\makebox(0,0)[lb]{$c_1$}}%
\put(32.0000,-3.0300){\makebox(0,0)[lb]{$y(t)$}}%
\put(7.4000,-3.7500){\makebox(0,0){$H_1(s)$}}%
%
\special{pn 8}%
\special{pa 1640 196}%
\special{pa 2000 196}%
\special{pa 2000 556}%
\special{pa 1640 556}%
\special{pa 1640 196}%
\special{fp}%
\put(18.2000,-3.7500){\makebox(0,0){$K$}}%
%
\special{pn 8}%
\special{pa 2000 358}%
\special{pa 2180 358}%
\special{dt 0.045}%
\special{sh 1}%
\special{pa 2180 358}%
\special{pa 2114 338}%
\special{pa 2128 358}%
\special{pa 2114 378}%
\special{pa 2180 358}%
\special{fp}%
%
\special{pn 8}%
\special{pa 1460 358}%
\special{pa 1640 358}%
\special{dt 0.045}%
\special{sh 1}%
\special{pa 1640 358}%
\special{pa 1574 338}%
\special{pa 1588 358}%
\special{pa 1574 378}%
\special{pa 1640 358}%
\special{fp}%
%
\special{pn 8}%
\special{pa 920 376}%
\special{pa 1100 376}%
\special{fp}%
\special{sh 1}%
\special{pa 1100 376}%
\special{pa 1034 356}%
\special{pa 1048 376}%
\special{pa 1034 396}%
\special{pa 1100 376}%
\special{fp}%
\put(20.2500,-3.0300){\makebox(0,0)[lb]{$c_2$}}%
%
\special{pn 8}%
\special{pa 2540 358}%
\special{pa 2720 358}%
\special{fp}%
\special{sh 1}%
\special{pa 2720 358}%
\special{pa 2654 338}%
\special{pa 2668 358}%
\special{pa 2654 378}%
\special{pa 2720 358}%
\special{fp}%
%
\special{pn 8}%
\special{pa 2720 178}%
\special{pa 3080 178}%
\special{pa 3080 538}%
\special{pa 2720 538}%
\special{pa 2720 178}%
\special{fp}%
%
\special{pn 8}%
\special{pa 3080 358}%
\special{pa 3440 358}%
\special{fp}%
\special{sh 1}%
\special{pa 3440 358}%
\special{pa 3374 338}%
\special{pa 3388 358}%
\special{pa 3374 378}%
\special{pa 3440 358}%
\special{fp}%
\put(29.0000,-3.5700){\makebox(0,0){$H_2(s)$}}%
\end{picture}%
\end{center}
\caption{Signal Reconstruction; the signal $x\in L^2$ is filtered by an analog filter $H_1(s)$ and
sampled by the ideal sampler $\samp{~}$.
Then an analog signal $y$ is produced by the zero-order hold $\hold{}$ and an analog filter $H_2(s)$.}
\label{fig:yy_dsp_sys}
\end{figure}

\begin{figure}[tb]
\begin{center}
\unitlength 0.1in
\begin{picture}( 32.5500,  8.4000)(  4.0000, -8.4000)
%
\special{pn 8}%
\special{pa 400 420}%
\special{pa 680 420}%
\special{fp}%
\special{sh 1}%
\special{pa 680 420}%
\special{pa 614 400}%
\special{pa 628 420}%
\special{pa 614 440}%
\special{pa 680 420}%
\special{fp}%
%
\special{pn 8}%
\special{pa 680 280}%
\special{pa 960 280}%
\special{pa 960 560}%
\special{pa 680 560}%
\special{pa 680 280}%
\special{fp}%
%
\special{pn 8}%
\special{pa 960 420}%
\special{pa 960 420}%
\special{fp}%
\special{pa 1100 420}%
\special{pa 1100 420}%
\special{fp}%
%
\special{pn 8}%
\special{pa 960 420}%
\special{pa 1100 420}%
\special{fp}%
%
\special{pn 8}%
\special{pa 1100 140}%
\special{pa 1100 700}%
\special{fp}%
%
\special{pn 8}%
\special{pa 1100 700}%
\special{pa 1240 700}%
\special{fp}%
\special{sh 1}%
\special{pa 1240 700}%
\special{pa 1174 680}%
\special{pa 1188 700}%
\special{pa 1174 720}%
\special{pa 1240 700}%
\special{fp}%
%
\special{pn 8}%
\special{pa 1240 560}%
\special{pa 1520 560}%
\special{pa 1520 840}%
\special{pa 1240 840}%
\special{pa 1240 560}%
\special{fp}%
%
\special{pn 8}%
\special{pa 1520 700}%
\special{pa 1660 700}%
\special{fp}%
\special{sh 1}%
\special{pa 1660 700}%
\special{pa 1594 680}%
\special{pa 1608 700}%
\special{pa 1594 720}%
\special{pa 1660 700}%
\special{fp}%
%
\special{pn 8}%
\special{pa 1660 560}%
\special{pa 1940 560}%
\special{pa 1940 840}%
\special{pa 1660 840}%
\special{pa 1660 560}%
\special{fp}%
%
\special{pn 8}%
\special{pa 2080 560}%
\special{pa 2360 560}%
\special{pa 2360 840}%
\special{pa 2080 840}%
\special{pa 2080 560}%
\special{fp}%
%
\special{pn 8}%
\special{pa 2500 560}%
\special{pa 2780 560}%
\special{pa 2780 840}%
\special{pa 2500 840}%
\special{pa 2500 560}%
\special{fp}%
%
\special{pn 8}%
\special{pa 2920 560}%
\special{pa 3200 560}%
\special{pa 3200 840}%
\special{pa 2920 840}%
\special{pa 2920 560}%
\special{fp}%
%
\special{pn 8}%
\special{pa 3200 700}%
\special{pa 3340 700}%
\special{fp}%
%
\special{pn 8}%
\special{pa 2780 700}%
\special{pa 2920 700}%
\special{fp}%
\special{sh 1}%
\special{pa 2920 700}%
\special{pa 2854 680}%
\special{pa 2868 700}%
\special{pa 2854 720}%
\special{pa 2920 700}%
\special{fp}%
%
\special{pn 8}%
\special{pa 3340 700}%
\special{pa 3340 456}%
\special{fp}%
\special{sh 1}%
\special{pa 3340 456}%
\special{pa 3320 522}%
\special{pa 3340 508}%
\special{pa 3360 522}%
\special{pa 3340 456}%
\special{fp}%
%
\special{pn 8}%
\special{pa 1940 0}%
\special{pa 2500 0}%
\special{pa 2500 280}%
\special{pa 1940 280}%
\special{pa 1940 0}%
\special{fp}%
%
\special{pn 8}%
\special{pa 1100 140}%
\special{pa 1940 140}%
\special{fp}%
\special{sh 1}%
\special{pa 1940 140}%
\special{pa 1874 120}%
\special{pa 1888 140}%
\special{pa 1874 160}%
\special{pa 1940 140}%
\special{fp}%
%
\special{pn 8}%
\special{pa 2500 140}%
\special{pa 3340 140}%
\special{fp}%
%
\special{pn 8}%
\special{pa 3340 140}%
\special{pa 3340 386}%
\special{fp}%
\special{sh 1}%
\special{pa 3340 386}%
\special{pa 3360 318}%
\special{pa 3340 332}%
\special{pa 3320 318}%
\special{pa 3340 386}%
\special{fp}%
%
\special{pn 8}%
\special{ar 3340 420 36 36  0.0000000 6.2831853}%
%
\special{pn 8}%
\special{pa 3376 420}%
\special{pa 3656 420}%
\special{fp}%
\special{sh 1}%
\special{pa 3656 420}%
\special{pa 3588 400}%
\special{pa 3602 420}%
\special{pa 3588 440}%
\special{pa 3656 420}%
\special{fp}%
%
\special{pn 8}%
\special{pa 1940 700}%
\special{pa 2080 700}%
\special{dt 0.045}%
\special{sh 1}%
\special{pa 2080 700}%
\special{pa 2014 680}%
\special{pa 2028 700}%
\special{pa 2014 720}%
\special{pa 2080 700}%
\special{fp}%
%
\special{pn 8}%
\special{pa 2360 700}%
\special{pa 2500 700}%
\special{dt 0.045}%
\special{sh 1}%
\special{pa 2500 700}%
\special{pa 2434 680}%
\special{pa 2448 700}%
\special{pa 2434 720}%
\special{pa 2500 700}%
\special{fp}%
\put(4.0000,-3.5500){\makebox(0,0)[lb]{$w(t)$}}%
\put(11.3500,-5.0000){\makebox(0,0)[lb]{$x(t)$}}%
\put(33.7500,-6.9500){\makebox(0,0)[lb]{$y(t)$}}%
\put(32.0000,-5.5500){\makebox(0,0)[lb]{$-$}}%
\put(28.7500,-1.1500){\makebox(0,0)[lb]{$x(t-L)$}}%
\put(34.5500,-3.5500){\makebox(0,0)[lb]{$e(t)$}}%
\put(22.2000,-1.4000){\makebox(0,0){$\ee^{-Ls}$}}%
\put(8.2000,-4.2000){\makebox(0,0){$F$}}%
\put(13.8000,-7.0000){\makebox(0,0){$H_1$}}%
\put(18.0000,-7.0000){\makebox(0,0){$\samp{}$}}%
\put(22.2000,-7.0000){\makebox(0,0){$K$}}%
\put(26.4000,-7.0000){\makebox(0,0){$\hold{}$}}%
\put(30.6000,-7.0000){\makebox(0,0){$H_2$}}%
\end{picture}%
\end{center}
\caption{Error system $\Err(K)$}
\label{fig:error_sys}
\end{figure}
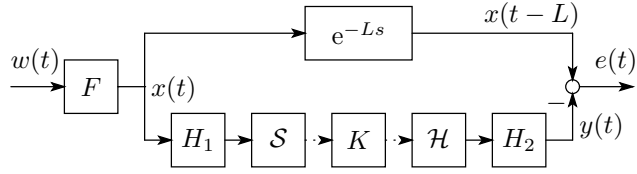

To optimize for the worst case, we consider the following
performance index:
\begin{equation}
J(K) = \sup_{\substack{x\in FL^2\\x\neq 0}}
\frac{
\left\|\left(\ee^{-Ls}-H_2\hold{~}K\samp{~}H_1\right)x\right\|_{L^2(\Rp)}}
{\|x\|_{L^2(\Rp)}}.
\label{eq:error}
\end{equation}
This is equivalent to the $H^\infty$ norm of the sampled-data error
system
\begin{equation}
 \Err(K):= (\ee^{-Ls}-H_2\hold{~}K\samp{~}H_1)F.
 \label{eq:error_sys}
\end{equation}
The block diagram of this error system is shown in Fig.\ \ref{fig:error_sys}.

\subsection{Computation of YY filter}
The optimal filter $\Kopt$ which minimizes $J(K)$ in (\ref{eq:error}) can be obtained
by numerical computation.
To compute the optimal filter $\Kopt$, we discretize the sampled-data error system $\Err(K)$
in (\ref{eq:error_sys}) by approximation \cite{KelAnd92,YamMadAnd99} or $H^\infty$ discretization \cite{MirTad03}.
We here discuss the approximation technique for minimizing $J(K)$ in (\ref{eq:error}).
We first introduce fast sampling and fast hold.
Let $\samp{N}$ and $\hold{N}$ are respectively
the ideal sampler and the zero-order hold with period $h/N$,
where $N$ is a positive integer ($N\geq 2$).
Then the system $\samp{N}\Err(K)\hold{N}$ becomes
a discrete-time multi-rate system with sampling periods
$h$ and $h/N$.
Then we introduce the {\it blocking operator} $\dlift{N}$,
or the {\it discrete-time lifting operator} \cite{CheFra,Nag03}:
\[
 \dlift{N}: \left\{v[0],v[1],v[2],\ldots\right\} \mapsto
 \left\{
	\begin{bmatrix}v[0]\\ v[1]\\ \vdots \\ v[N-1]\end{bmatrix},
	\begin{bmatrix}v[N]\\ v[N+1]\\ \vdots \\ v[2N-1]\end{bmatrix},
	\ldots
 \right\}.
\]
This operator converts a 1-dimensional signal $v$ into an $N$-dimensional signal
and the sampling rate becomes $N$ times slower.
This operation makes it possible to equivalently convert multirate systems into 
single-rate ones, and hence the analysis and design become easier.
By using this operator,
the system $E_N(K)$ defined by
\begin{equation}
 E_N(K):= \dlift{N} \samp{N} \Err(K) \hold{N} \idlift{N}
 \label{eq:error_sys_apr}
\end{equation}
becomes a discrete-time LTI system.
Moreover, we can say that
for any integer $N\geq 2$ and any stable $K$, there exist discrete-time LTI systems
$G_{1,N}$, $G_{2,N}$ and $G_{3,N}$ such that \cite{Nag03}
\[
 E_N(K) = G_{1,N} + G_{2,N}KG_{3,N},
\]
and the LTI system $E_N(K)$ is approximation of $\Err(K)$ in the sense that \cite{YamMadAnd99}
\[
 \lim_{N\rightarrow\infty} \|E_N(K)\|_\infty \rightarrow J(K) = \|\Err(K)\|_\infty.
\]
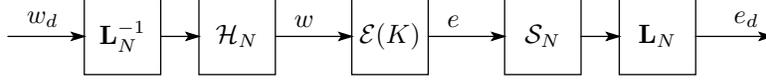
\begin{figure}[tb]
\begin{center}
\unitlength 0.1in
\begin{picture}( 40.0000,  4.2000)(  4.0000, -6.0000)
%
\special{pn 8}%
\special{pa 400 400}%
\special{pa 800 400}%
\special{fp}%
\special{sh 1}%
\special{pa 800 400}%
\special{pa 734 380}%
\special{pa 748 400}%
\special{pa 734 420}%
\special{pa 800 400}%
\special{fp}%
%
\special{pn 8}%
\special{pa 800 200}%
\special{pa 1200 200}%
\special{pa 1200 600}%
\special{pa 800 600}%
\special{pa 800 200}%
\special{fp}%
%
\special{pn 8}%
\special{pa 1200 400}%
\special{pa 1400 400}%
\special{fp}%
\special{sh 1}%
\special{pa 1400 400}%
\special{pa 1334 380}%
\special{pa 1348 400}%
\special{pa 1334 420}%
\special{pa 1400 400}%
\special{fp}%
%
\special{pn 8}%
\special{pa 1400 200}%
\special{pa 1800 200}%
\special{pa 1800 600}%
\special{pa 1400 600}%
\special{pa 1400 200}%
\special{fp}%
%
\special{pn 8}%
\special{pa 1800 400}%
\special{pa 2200 400}%
\special{fp}%
\special{sh 1}%
\special{pa 2200 400}%
\special{pa 2134 380}%
\special{pa 2148 400}%
\special{pa 2134 420}%
\special{pa 2200 400}%
\special{fp}%
%
\special{pn 8}%
\special{pa 2200 200}%
\special{pa 2600 200}%
\special{pa 2600 600}%
\special{pa 2200 600}%
\special{pa 2200 200}%
\special{fp}%
%
\special{pn 8}%
\special{pa 2600 400}%
\special{pa 3000 400}%
\special{fp}%
\special{sh 1}%
\special{pa 3000 400}%
\special{pa 2934 380}%
\special{pa 2948 400}%
\special{pa 2934 420}%
\special{pa 3000 400}%
\special{fp}%
%
\special{pn 8}%
\special{pa 3000 200}%
\special{pa 3400 200}%
\special{pa 3400 600}%
\special{pa 3000 600}%
\special{pa 3000 200}%
\special{fp}%
%
\special{pn 8}%
\special{pa 3400 400}%
\special{pa 3600 400}%
\special{fp}%
\special{sh 1}%
\special{pa 3600 400}%
\special{pa 3534 380}%
\special{pa 3548 400}%
\special{pa 3534 420}%
\special{pa 3600 400}%
\special{fp}%
%
\special{pn 8}%
\special{pa 3600 200}%
\special{pa 4000 200}%
\special{pa 4000 600}%
\special{pa 3600 600}%
\special{pa 3600 200}%
\special{fp}%
%
\special{pn 8}%
\special{pa 4000 400}%
\special{pa 4400 400}%
\special{fp}%
\special{sh 1}%
\special{pa 4400 400}%
\special{pa 4334 380}%
\special{pa 4348 400}%
\special{pa 4334 420}%
\special{pa 4400 400}%
\special{fp}%
\put(24.0000,-4.0000){\makebox(0,0){$\Err(K)$}}%
\put(16.0000,-4.0000){\makebox(0,0){$\hold{N}$}}%
\put(10.0000,-4.0000){\makebox(0,0){$\idlift{N}$}}%
\put(32.0000,-4.0000){\makebox(0,0){$\samp{N}$}}%
\put(38.0000,-4.0000){\makebox(0,0){$\dlift{N}$}}%
\put(19.0000,-3.5000){\makebox(0,0)[lb]{$w$}}%
\put(27.0000,-3.5000){\makebox(0,0)[lb]{$e$}}%
\put(42.0000,-3.5000){\makebox(0,0)[lb]{$e_d$}}%
\put(5.0000,-3.5000){\makebox(0,0)[lb]{$w_d$}}%
\end{picture}%
\end{center}
\caption{Fast discretization of the sampled-data system $\Err(K)$: $\dlift{N}$ is the blocking operator,
$\samp{N}$ and $\hold{N}$ are respectively the fast sampler and the fast hold with sampling period $h/N$.}
\label{fig:fsfh}
\end{figure}
The optimization of minimizing $E_N(K)$ is easily done by using discrete-time $H^\infty$
optimization technique.
We can therefore obtain a stable and causal filter $K$ which approximates the optimal filter $\Kopt$.

\subsection{Robustness}
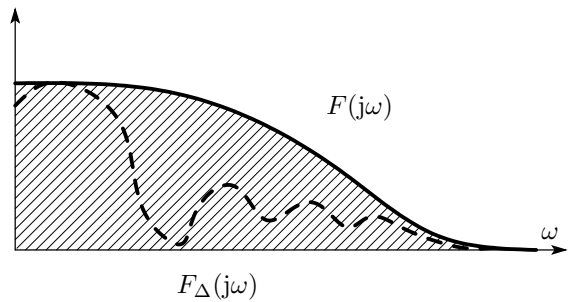
\begin{figure}[tb]
\begin{center}
\unitlength 0.1in
\begin{picture}(28.85,13.51)(3.95,-15.51)
%
\special{pn 8}%
\special{pa 400 1460}%
\special{pa 400 200}%
\special{fp}%
\special{sh 1}%
\special{pa 400 200}%
\special{pa 380 267}%
\special{pa 400 253}%
\special{pa 420 267}%
\special{pa 400 200}%
\special{fp}%
%
\special{pn 8}%
\special{pa 400 1460}%
\special{pa 3280 1460}%
\special{fp}%
\special{sh 1}%
\special{pa 3280 1460}%
\special{pa 3213 1440}%
\special{pa 3227 1460}%
\special{pa 3213 1480}%
\special{pa 3280 1460}%
\special{fp}%
%
\special{pn 20}%
\special{pa 400 587}%
\special{pa 432 587}%
\special{pa 465 586}%
\special{pa 497 586}%
\special{pa 530 586}%
\special{pa 562 586}%
\special{pa 594 586}%
\special{pa 627 586}%
\special{pa 659 586}%
\special{pa 691 586}%
\special{pa 724 587}%
\special{pa 756 588}%
\special{pa 788 589}%
\special{pa 820 590}%
\special{pa 852 591}%
\special{pa 884 593}%
\special{pa 916 595}%
\special{pa 948 598}%
\special{pa 980 600}%
\special{pa 1011 604}%
\special{pa 1043 607}%
\special{pa 1075 611}%
\special{pa 1106 615}%
\special{pa 1137 620}%
\special{pa 1169 626}%
\special{pa 1200 631}%
\special{pa 1231 638}%
\special{pa 1262 645}%
\special{pa 1293 652}%
\special{pa 1323 660}%
\special{pa 1354 668}%
\special{pa 1384 677}%
\special{pa 1415 687}%
\special{pa 1445 697}%
\special{pa 1475 707}%
\special{pa 1505 718}%
\special{pa 1535 729}%
\special{pa 1564 741}%
\special{pa 1594 754}%
\special{pa 1623 766}%
\special{pa 1652 779}%
\special{pa 1682 793}%
\special{pa 1711 807}%
\special{pa 1739 822}%
\special{pa 1768 836}%
\special{pa 1797 852}%
\special{pa 1825 867}%
\special{pa 1853 883}%
\special{pa 1881 900}%
\special{pa 1909 917}%
\special{pa 1937 934}%
\special{pa 1964 951}%
\special{pa 1992 969}%
\special{pa 2019 987}%
\special{pa 2046 1006}%
\special{pa 2073 1024}%
\special{pa 2100 1043}%
\special{pa 2126 1063}%
\special{pa 2153 1082}%
\special{pa 2179 1102}%
\special{pa 2205 1122}%
\special{pa 2231 1142}%
\special{pa 2257 1162}%
\special{pa 2283 1182}%
\special{pa 2309 1202}%
\special{pa 2335 1222}%
\special{pa 2362 1241}%
\special{pa 2388 1259}%
\special{pa 2414 1278}%
\special{pa 2441 1295}%
\special{pa 2468 1312}%
\special{pa 2495 1328}%
\special{pa 2522 1344}%
\special{pa 2550 1358}%
\special{pa 2578 1372}%
\special{pa 2607 1384}%
\special{pa 2636 1395}%
\special{pa 2665 1405}%
\special{pa 2695 1414}%
\special{pa 2726 1422}%
\special{pa 2756 1429}%
\special{pa 2787 1434}%
\special{pa 2819 1439}%
\special{pa 2851 1444}%
\special{pa 2882 1447}%
\special{pa 2915 1450}%
\special{pa 2947 1453}%
\special{pa 2980 1455}%
\special{pa 3012 1456}%
\special{pa 3045 1457}%
\special{pa 3078 1459}%
\special{pa 3111 1460}%
\special{pa 3127 1460}%
\special{sp}%
%
\special{pn 4}%
\special{pa 1741 821}%
\special{pa 1102 1460}%
\special{fp}%
\special{pa 1705 803}%
\special{pa 1048 1460}%
\special{fp}%
\special{pa 1669 785}%
\special{pa 994 1460}%
\special{fp}%
\special{pa 1633 767}%
\special{pa 940 1460}%
\special{fp}%
\special{pa 1597 749}%
\special{pa 886 1460}%
\special{fp}%
\special{pa 1552 740}%
\special{pa 832 1460}%
\special{fp}%
\special{pa 1516 722}%
\special{pa 778 1460}%
\special{fp}%
\special{pa 1471 713}%
\special{pa 724 1460}%
\special{fp}%
\special{pa 1435 695}%
\special{pa 670 1460}%
\special{fp}%
\special{pa 1390 686}%
\special{pa 616 1460}%
\special{fp}%
\special{pa 1354 668}%
\special{pa 562 1460}%
\special{fp}%
\special{pa 1309 659}%
\special{pa 508 1460}%
\special{fp}%
\special{pa 1264 650}%
\special{pa 454 1460}%
\special{fp}%
\special{pa 1228 632}%
\special{pa 409 1451}%
\special{fp}%
\special{pa 1174 632}%
\special{pa 400 1406}%
\special{fp}%
\special{pa 1129 623}%
\special{pa 400 1352}%
\special{fp}%
\special{pa 1084 614}%
\special{pa 400 1298}%
\special{fp}%
\special{pa 1039 605}%
\special{pa 400 1244}%
\special{fp}%
\special{pa 985 605}%
\special{pa 400 1190}%
\special{fp}%
\special{pa 940 596}%
\special{pa 400 1136}%
\special{fp}%
\special{pa 886 596}%
\special{pa 400 1082}%
\special{fp}%
\special{pa 841 587}%
\special{pa 400 1028}%
\special{fp}%
\special{pa 787 587}%
\special{pa 400 974}%
\special{fp}%
\special{pa 733 587}%
\special{pa 400 920}%
\special{fp}%
\special{pa 679 587}%
\special{pa 400 866}%
\special{fp}%
\special{pa 625 587}%
\special{pa 400 812}%
\special{fp}%
\special{pa 571 587}%
\special{pa 400 758}%
\special{fp}%
\special{pa 517 587}%
\special{pa 400 704}%
\special{fp}%
\special{pa 463 587}%
\special{pa 400 650}%
\special{fp}%
\special{pa 1777 839}%
\special{pa 1156 1460}%
\special{fp}%
%
\special{pn 4}%
\special{pa 1813 857}%
\special{pa 1210 1460}%
\special{fp}%
\special{pa 1840 884}%
\special{pa 1264 1460}%
\special{fp}%
\special{pa 1876 902}%
\special{pa 1318 1460}%
\special{fp}%
\special{pa 1912 920}%
\special{pa 1372 1460}%
\special{fp}%
\special{pa 1948 938}%
\special{pa 1426 1460}%
\special{fp}%
\special{pa 1975 965}%
\special{pa 1480 1460}%
\special{fp}%
\special{pa 2011 983}%
\special{pa 1534 1460}%
\special{fp}%
\special{pa 2047 1001}%
\special{pa 1588 1460}%
\special{fp}%
\special{pa 2074 1028}%
\special{pa 1642 1460}%
\special{fp}%
\special{pa 2101 1055}%
\special{pa 1696 1460}%
\special{fp}%
\special{pa 2137 1073}%
\special{pa 1750 1460}%
\special{fp}%
\special{pa 2173 1091}%
\special{pa 1804 1460}%
\special{fp}%
\special{pa 2200 1118}%
\special{pa 1858 1460}%
\special{fp}%
\special{pa 2227 1145}%
\special{pa 1912 1460}%
\special{fp}%
\special{pa 2254 1172}%
\special{pa 1966 1460}%
\special{fp}%
\special{pa 2290 1190}%
\special{pa 2020 1460}%
\special{fp}%
\special{pa 2317 1217}%
\special{pa 2074 1460}%
\special{fp}%
\special{pa 2353 1235}%
\special{pa 2128 1460}%
\special{fp}%
\special{pa 2380 1262}%
\special{pa 2182 1460}%
\special{fp}%
\special{pa 2416 1280}%
\special{pa 2236 1460}%
\special{fp}%
\special{pa 2443 1307}%
\special{pa 2290 1460}%
\special{fp}%
\special{pa 2479 1325}%
\special{pa 2344 1460}%
\special{fp}%
\special{pa 2515 1343}%
\special{pa 2398 1460}%
\special{fp}%
\special{pa 2551 1361}%
\special{pa 2452 1460}%
\special{fp}%
\special{pa 2587 1379}%
\special{pa 2506 1460}%
\special{fp}%
\special{pa 2623 1397}%
\special{pa 2560 1460}%
\special{fp}%
\special{pa 2668 1406}%
\special{pa 2614 1460}%
\special{fp}%
\special{pa 2713 1415}%
\special{pa 2668 1460}%
\special{fp}%
\special{pa 2749 1433}%
\special{pa 2722 1460}%
\special{fp}%
%
\special{pn 20}%
\special{pa 400 704}%
\special{pa 424 680}%
\special{pa 449 656}%
\special{pa 474 635}%
\special{pa 499 616}%
\special{pa 526 600}%
\special{pa 555 589}%
\special{pa 585 583}%
\special{pa 616 582}%
\special{pa 649 585}%
\special{pa 681 591}%
\special{pa 713 601}%
\special{pa 745 613}%
\special{pa 775 628}%
\special{pa 804 644}%
\special{pa 831 663}%
\special{pa 857 683}%
\special{pa 881 705}%
\special{pa 903 728}%
\special{pa 923 753}%
\special{pa 941 780}%
\special{pa 956 807}%
\special{pa 969 836}%
\special{pa 980 866}%
\special{pa 989 896}%
\special{pa 997 927}%
\special{pa 1004 959}%
\special{pa 1009 991}%
\special{pa 1015 1023}%
\special{pa 1020 1055}%
\special{pa 1026 1087}%
\special{pa 1031 1119}%
\special{pa 1038 1150}%
\special{pa 1046 1181}%
\special{pa 1055 1211}%
\special{pa 1066 1241}%
\special{pa 1079 1269}%
\special{pa 1095 1297}%
\special{pa 1113 1324}%
\special{pa 1134 1350}%
\special{pa 1159 1375}%
\special{pa 1187 1398}%
\special{pa 1217 1419}%
\special{pa 1246 1431}%
\special{pa 1270 1431}%
\special{pa 1288 1417}%
\special{pa 1301 1391}%
\special{pa 1311 1357}%
\special{pa 1322 1319}%
\special{pa 1336 1280}%
\special{pa 1354 1242}%
\special{pa 1377 1208}%
\special{pa 1402 1177}%
\special{pa 1430 1152}%
\special{pa 1459 1134}%
\special{pa 1489 1122}%
\special{pa 1518 1118}%
\special{pa 1545 1124}%
\special{pa 1570 1139}%
\special{pa 1594 1163}%
\special{pa 1615 1191}%
\special{pa 1636 1222}%
\special{pa 1656 1252}%
\special{pa 1677 1278}%
\special{pa 1699 1297}%
\special{pa 1723 1307}%
\special{pa 1748 1305}%
\special{pa 1775 1294}%
\special{pa 1804 1276}%
\special{pa 1834 1255}%
\special{pa 1864 1235}%
\special{pa 1895 1219}%
\special{pa 1926 1209}%
\special{pa 1956 1210}%
\special{pa 1985 1220}%
\special{pa 2013 1239}%
\special{pa 2040 1264}%
\special{pa 2067 1291}%
\special{pa 2092 1316}%
\special{pa 2118 1335}%
\special{pa 2144 1344}%
\special{pa 2170 1338}%
\special{pa 2197 1320}%
\special{pa 2225 1300}%
\special{pa 2254 1286}%
\special{pa 2284 1283}%
\special{pa 2315 1289}%
\special{pa 2346 1302}%
\special{pa 2377 1318}%
\special{pa 2408 1335}%
\special{pa 2438 1350}%
\special{pa 2468 1364}%
\special{pa 2498 1377}%
\special{pa 2527 1388}%
\special{pa 2556 1399}%
\special{pa 2586 1409}%
\special{pa 2616 1418}%
\special{pa 2646 1428}%
\special{pa 2677 1436}%
\special{pa 2709 1443}%
\special{pa 2740 1449}%
\special{pa 2772 1452}%
\special{pa 2804 1454}%
\special{pa 2836 1454}%
\special{pa 2868 1452}%
\special{pa 2900 1449}%
\special{pa 2932 1446}%
\special{pa 2964 1442}%
\special{pa 2965 1442}%
\special{sp 0.070}%
\put(20.2900,-7.8500){\makebox(0,0)[lb]{$F(\jj\omega)$}}%
\put(12.4600,-17.2100){\makebox(0,0)[lb]{$F_\Delta(\jj\omega)$}}%
\put(31.5400,-14.0600){\makebox(0,0)[lb]{$\omega$}}%
\end{picture}%
\end{center}
\caption{Nominal filter $F(s)$ (solid) and perturbed $F_\Delta(s)$ (dash)}
\label{fig:robust}
\end{figure}

Next let us consider robustness against uncertainty of the analog signal model $F(s)$.
In practice, $F(s)$ cannot be identified exactly.
We therefore partially circumvent this defect
by discussing the robustness of the filter against uncertainty of $F(s)$. 
Let us assume the unstructured uncertainty of
the following type: 
\begin{gather*}
F_{\Delta}(s) := F(s)(1+\Delta(s)),\quad
\Err^\Delta(K) := \left(\ee^{-Ls} - H_2\hold{~}K\samp{~}H_1\right)F_{\Delta},\\
\Delta \in {\boldsymbol \Delta}:=\left\{\Delta : \|1+\Delta\|_\infty\leq \gamma \right\}.
\end{gather*}
Then we have the following proposition: 
\begin{prop} 
For any stable $K$ and $\Delta \in {\boldsymbol \Delta}$,
we have $\|\Err^\Delta(K)\|_\infty \leq \gamma \|\Err(K)\|_\infty$.
\end{prop}
By this proposition, the nominal performance $\|\Err(K)\|_\infty$ is guaranteed
against the perturbation $\Delta\in {\boldsymbol \Delta}$ if $\gamma \leq 1$.
In some cases, it is possible that $\gamma = 1$, in which
case the performance is bounded as illustrated in 
Fig.\  \ref{fig:robust}.  This means that if we take 
$F(s)$ that covers all possible gain characteristics of the
input analog signals, it gives a bound for the error norm.
This at least partially justifies the choice of the
first-order weighting $F(s)$ in the previous section.  

\subsection{FIR YY filter by LMI}
The error system (\ref{eq:error_sys}) or (\ref{eq:error_sys_apr}
is {\it affine} in the filter $K$ to be designed.
By this fact, we can design the optimal FIR (finite impulse response) filter of the form
\[
 K(z) = \sum_{n=0}^N a_nz^{-n}.
\]
By this, the error system (\ref{eq:error_sys_apr}) is affine
in the design parameter $a_0,a_1,\ldots,a_N$.
It follows that the optimization of minimizing $\|E_N(K)\|_\infty$
can be described by an LMI (linear matrix inequality)
by the bounded real lemma or Kalman-Yakubovic-Popov lemma \cite{YamAndNagKoy03}.
The optimization with an LMI can be solved easily by computer softwares.

\subsection{Summary}
The advantages of the YY filter discussed in this section
are the following:
\begin{enumerate}
\item the optimal filter is always causal and stable.
\item the design takes the inter-sample behavior into account.
\item the system is robust against the uncertainty of input signals.
\item the optimal FIR filter is also obtainable
via an LMI.
\end{enumerate}

\section{Conclusions}
In this article, problems in Shannon's theorem have been pointed out
and the advantages of YY filter over the conventional signal processing
have been introduced.
In fact, YY filters are implemented in 
commercial MD players, silicon-audio devices,
and  mobile phones.
One of future works is
design of adaptive YY filters.


\end{document}